\def\year{2019}\relax
\documentclass[letterpaper]{article} 
\usepackage{aaai19}  
\usepackage{times}  
\usepackage{helvet}  
\usepackage{courier}  
\usepackage{url}  
\usepackage{graphicx}  

\usepackage{amsmath}
\usepackage{upgreek}
\usepackage{multirow}
\usepackage{pdfpages}
\usepackage{array}

\title{A Domain Generalization Perspective on Listwise Context Modeling}

\author{Lin Zhu, Yihong Chen, Bowen He\\
	Ctrip Travel Network Technology Co., Limited.\\
	\{zhulb, yihongchen, bwhe\}@ctrip.com\\
}

\begin{document}

\maketitle
\begin{abstract}
As one of the most popular techniques for solving the ranking problem in information retrieval, Learning-to-rank (LETOR) has received a lot of attention both in academia and industry due to its importance in a wide variety of data mining applications. However, most of existing LETOR approaches choose to learn a single global ranking function to handle all queries, and ignore the substantial differences that exist between queries. In this paper, we propose a domain generalization strategy to tackle this problem. We propose Query-Invariant Listwise Context Modeling (QILCM), a novel neural architecture which eliminates the detrimental influence of inter-query variability by learning \textit{query-invariant} latent representations, such that the ranking system could generalize better to unseen queries. We evaluate our techniques on benchmark datasets, demonstrating that QILCM outperforms previous state-of-the-art approaches by a substantial margin.
\end{abstract}

\section{Introduction}

As an important learning paradigm for tackling the ranking problem in information retrieval, learning-to-rank (LETOR) has received a lot of attention both in academia and industry due to its importance in a wide variety of applications, such as document retrieval \cite{ai2018learning,YahooL2R}, recommendation systems \cite{freno2017practical,volkovs2012collaborative}, and E-commerce search \cite{zhuang2018globally,karmaker2017application}. 

The typical assumption behind LETOR methods is that for any given query, the relevance score of a candidate item is a parameterized \textit{global} function of a set of features that describe the query, the item, and their interactions \cite{karmaker2017application,ai2018learning}. The parameters of this function can be determined by fitting it to a training set that consists of query-item pairs and the associated relevance judgments. Once learned, the function is applied to handle any future query to rank the candidate items with respect the query. However, queries may vary significantly in the underlying users' intentions \cite{geng2008query}, in the distributions of relevant items \cite{ai2018learning}, or in the features that are relevant to item ranking \cite{karmaker2017application}, and it is difficult to learn a single model that could encompass all these diversities. Moreover, given the substantial variations between queries, it is quite possible that a unseen test query have certain characteristics that are rarely encountered in the training set, leading to degradation in generalization ability of the ranking function \cite{karmaker2017application}. 

So far, a number of approaches have been proposed to tackle the aforementioned problem. In particular, it has been demonstrated that the performance of a global ranking model can be greatly improved by incorporating the \textit{local ranking context} information \cite{ai2018learning}, which is generally extracted from the input data via feature engineering efforts or novel model architectures \cite{geng2008query,ai2018learning,DCM,zhuang2018globally}. Although some promising results have been shown,  these approaches still do not explicitly filter out the detrimental variations of queries, which may increase the risk of overfitting and hurt the performance.

At another extreme, we could train a ranking model for every query independently \cite{ai2018learning,geng2008query}. While this strategy may avoid the negative influence of the heterogeneity of queries, it does not allow information to be shared and transfered between queries, which makes generalization to unseen queries even harder.

A better solution may lie in the middle of these two extremes: we still learn a common ranking model for all queries, and yet treat every query as a separate \textit{domain}. All of the queries (domains) share the same learning task and the same input feature space, but have different distributions. In recognition of the variations between queries (domains), we would like the model to acquire knowledge from `source' queries (domains) that are available as the training set, and adapt the learned knowledge to unknown `target' queries (domains).

In this formulation, we notice that the LETOR problem is equivalent to a \textit{domain generalization (DG)} problem \cite{blanchard2011generalizing}, where the objective is precisely to infer a learning system that can take as input a set of training domains and will output a model that can achieve high accuracy for new domains. To achieve this goal, many existing DG methods propose to learn \textit{domain-invariant} representations that are expected to remove the negative effects of distributional changes across domains \cite{muandet2013domain,xie2017controllable}. 

On the basis of the above considerations, in this paper we propose to solve the query heterogeneity problem via DG techniques. The main contributions of our work are summarized as follows:
\begin{itemize}
	\item We introduce a DG formulation of the LETOR problem and discuss why a DG perspective is justified in this setting;
	\item We propose Query-Invariant Listwise Context Model (QILCM), a novel neural architecture for DG in this LETOR context;
	\item We perform comprehensive evaluations on three benchmark datasets, demonstrating that QILCM outperforms previous state-of-the-art approaches by a substantial margin.	
\end{itemize}

\section{Related Work}

\subsection{LETOR methods}

According to how training losses are formulated, existing algorithms for LETOR can be broadly categorized into three categories: (1) The pointwise approaches \cite{li2008mcrank,nallapati2004discriminative}, which reformulates LETOR as a classification or regression problem on single items;(2) The pairwise approaches \cite{freund2003efficient,burges2005learning}, which formulate LETOR as a classification problem on the item pairs; (3) The listwise approaches \cite{SoftRank,000278621400005} which directly optimize the ranking metrics. As mentioned in the introduction section, queries could vary greatly in various aspects, and exhibit different `standards' for ranking \cite{sqlrank}. As a result, listwise and pairwise LETOR methods, which mainly focus on modeling the relative (instead of absolute) preferences between items for a query, generally perform better than pointwise approaches in practice \cite{karmaker2017application,cao2007learning,YahooL2R}. Our proposed DG perspective for LETOR could be considered as a continuation of these methods, since it more explicitly enforces the model to filter out the `global' differences between queries, and focus on the `relative' differences between items for each given query. Meanwhile, most of existing listwise and pairwise LETOR methods still try to learn a global ranking function that assesses the relevance of each item individually. In contrast, listwise context modeling methods studied in this paper treat the candidate items as a whole, and can better model their mutual influences \cite{zhuang2018globally}.

\subsection{Neural Ranking Models}

Deep neural network methods have been applied to numerous ranking applications, such as recommendation systems \cite{covington2016deep}, ad-hoc retrieval \cite{fan2018modeling,zamani2018neural}, and context-aware ranking \cite{zamani2017situational}. Similar to the traditional LETOR methods discussed above, these models also score each item independently and neglect the negative influence of distributional differences between queries. This limitation is clearly demonstrated in \cite{cohen2018cross}, which shows that these models could overfit to the domains from which the training data is sampled, and generalize poorly to domains not encountered during training. To handle this problem, Cohen et al. also propose learning of domain invariant representations. However, they assume that the queries have already been categorized into a few broad domains and focus on avoidance of overfitting to these predetermined domains, while our method do not rely on such additional supervisory information, and instead focus directly on tackling the finer-grained variations between queries.

\subsection{Context Aware Ranking}

A significant amount of research has focused on leveraging contextual data to improve ranking. In
particular, certain types of contextual information have been explored in depth, such as temporal dynamics of user behaviors \cite{xiang2010context,chen2018sequential,ATRANK} and the geological location data \cite{zamani2017situational,manotumruksa2018contextual}. 

On the other hand, modeling of the local ranking context formed by candidate items is a less well studied problem, and state-of-the-art approaches are generally based on the Recurrent Neural Network (RNN) models \cite{DCM,ai2018learning,zhuang2018globally}. For example, \cite{zhuang2018globally} reformulate the ranking task as a RNN-based sequence generation problem and use the beam search algorithm to generate the ranked item list, while Deep Choice Model (DCM) \cite{DCM} and Deep Listwise Context Model (DLCM) \cite{ai2018learning} adopt RNNs to generate a context encoding that summarizes the candidate items, which is then adopted to re-query the candidate items to generate the ranking scores. Although some promising results have been shown, a potential limitation of these methods is that RNNs are mainly designed for modeling sequential data, and yet the data encountered in ranking problems cannot always be organized as a natural and unique sequence. For example, the candidate items may be selected from a large repository using some initial retrieval methods \cite{covington2016deep,YahooL2R}, and are essentially orderless. For such problems, although one can still order the candidate items according to certain pre-specified rules, and in principle universal approximators such as RNNs should be robust to choices of such pre-ordering, previous studies nevertheless show that the input order enforces a implicit prior on how RNNs should encode the given data, and could have strong impact on the experimental performance \cite{OrderMatters}. Meanwhile, by adopting additional components such as positional encoding \cite{vaswani2017attention}, models without recurrent structures can also incorporate useful order information to achieve good performance \cite{vaswani2017attention,gehring2017convolutional,ATRANK}.  

In addition, it is still necessary for these approaches to learn from the data to filter out the detrimental variations of queries, while preserving the necessary information for the ranking task. Better performance may be obtained by relieving such a learning burden using certain dedicated architectures that explicitly encode robustness to inter-query variations, such as the DG models reviewed later.

\subsection{Domain Generalization (DG)}

So far, a large number of formulations have been proposed to solve the DG problem, we refer the interested readers to \cite{li2018learning,shankar2018generalizing} for up-to-date reviews of related techniques. Our method is directly inspired by the DG approaches which learn a high-level data representation that simultaneously minimizes domain dissimilarity and preserves the functional relationship with the learning target. To the best of our knowledge, this is the first time that such techniques are applied to model the local ranking context formed by candidate items.

\section{Methodology}

\subsection{Problem Formulation}

Let $\mathcal{Q}$ be the set of possible queries and $\mathcal{D}$ be the set of possible items. Assume we are given a set of queries $\mathcal{T}\subseteq \mathcal{Q}$ as the training data, where each query $q\in \mathcal{T}$ is provided with a set of $n_{q}$ candidate items $\left\{ {{d}_{q,1}},{{d}_{q,2}},\cdots ,{{d}_{q,{{n}_{q}}}} \right\}\in \mathcal{D}$ to be ranked. Each pair of query-item $\left( q,{{d}_{q,i}} \right),1\le i\le {{n}_{q}}$ is described by the same set of categorical and numerical features, and assigned a score ${{y}_{q,i}}$ that quantifies the relevance of ${{d}_{q,i}}$ to $q$. The objective of LETOR is to infer a ranking model that can accurately assess the relevance of any given query-item pair. 

\subsection{Model Architecture}

\begin{figure*}
	\centering
	\includegraphics[width=1.5\columnwidth]{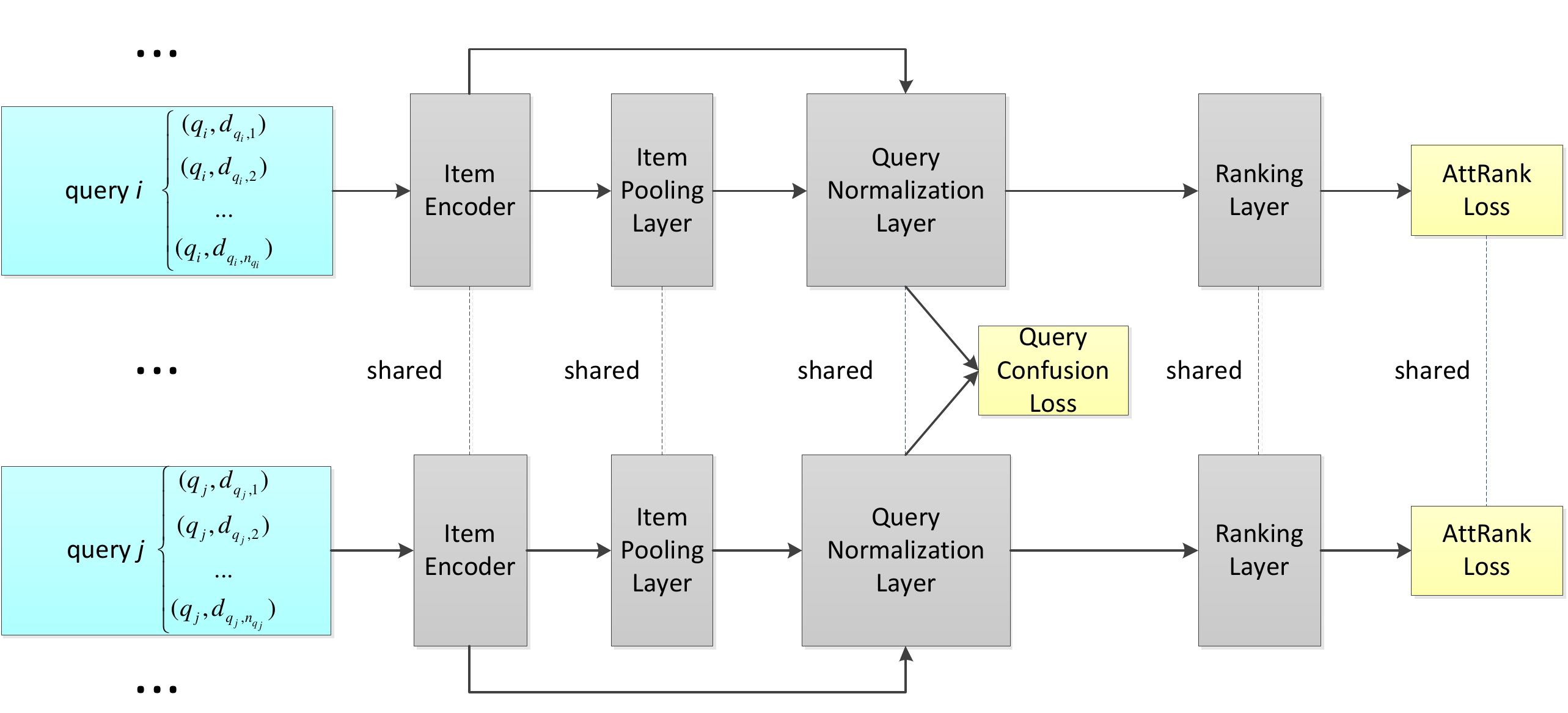} 
	\caption{The overall architecture of Query-Invariant Listwise Context Model (QILCM).}
	\label{Architecture}
\end{figure*}

As discussed in the introduction, we propose to cast LETOR as a DG problem, and view every query as an individual domain. All of the domains/queries share the same learning task and the same input feature space, but have different distributions. To better account for such distributional differences, we parametrize the desired ranking model as a neural architecture that consists of five main components, as illustrated in Figure \ref{Architecture}:

\begin{itemize}
	\item An \textbf{item encoder} that transforms each query-item pair $\left(q,{{d}_{q,i}} \right)$ into an encoding vector ${{\mathbf{h}}_{q,i}}$;
	\item An exchangeable \textbf{item pooling layer} that collapses the encoding vectors associated with each query $q$ to a single query embedding vector ${{\mathbf{c}}_{q}}$, which is then combined with ${{\mathbf{h}}_{q,i}}$ to construct a refined item encoding vector ${{\widetilde{\mathbf{h}}}_{q,i}}$;  
	\item A parameter-free \textbf{query normalization layer} that reduces the distributional differences between item encodings from different queries; 
	\item A \textbf{ranking layer} that computes the ranking score using item representations learned from the previous layers;
	\item An \textbf{objective function} that simultaneously promotes the improvement of ranking accuracy and the semantic alignment of feature distributions among queries.
\end{itemize}

\subsection{Item Encoder}

The item encoder that we consider in this paper is similar to previous works \cite{DCM,ai2018learning,covington2016deep}. It accepts the features of query-item pairs as input. Given the sensitivity of neural networks to input scaling \cite{lecun2012efficient,covington2016deep}, numerical features are normalized to the interval $\left[ 0,1 \right]$, and categorical features are mapped to dense vectors via embeddings, which are learned jointly with all other model parameters through back-propagation. 

Additionally, sometimes the input data may be supplemented with certain order information, such as the predictive results given by an initial LETOR algorithm \cite{ai2018learning}. Different from previous works that model such information using RNNs, we simply incorporate it as an additional numerical feature. While more sophisticated schemes for encoding order information \cite{vaswani2017attention,ATRANK,gehring2017convolutional} are readily applicable, we found that such a simple option already performs well on the datasets that we tested.

For each pair of $\left( q,{{d}_{q,i}} \right)$, all of the pre-processed features mentioned above are concatenated into an array ${\mathbf{x}}_{q,i}$. As noted in \cite{ai2018learning}, direct usage of this feature representation may fail to fully leverage the expressive power of neural models. We thus follow \cite{ai2018learning}, and pass ${\mathbf{x}}_{q,i}$ through 2 layers of fully connected exponential linear units (ELUs) \cite{elu}:
\begin{equation}
	\begin{matrix}
	\mathbf{x}_{q,i}^{(1)}=\text{ELU}\left( {{\mathbf{W}}^{(1)}}{{\mathbf{x}}_{q,i}}+{{\mathbf{b}}^{(1)}} \right), \\ 
	\mathbf{x}_{q,i}^{(2)}=\text{ELU}\left( {{\mathbf{W}}^{(2)}}\mathbf{x}_{q,i}^{(1)}+{{\mathbf{b}}^{(2)}} \right), \\ 
	\end{matrix}
	\label{abstract}
\end{equation}
the output of which is then concatenated with ${\mathbf{x}}_{q,i}$ to construct a new feature vector ${{\mathbf{h}}_{q,i}}$:
\begin{equation}
{{\mathbf{h}}_{q,i}}=\left[ \begin{matrix}
{{\mathbf{x}}_{q,i}}  \\
\mathbf{x}_{q,i}^{(2)}  \\
\end{matrix} \right].
\end{equation}

\subsection{Item Pooling Layer}

In this work, different queries are treated as different domains, since we observe only a subset of the possible queries/domains during training, additional assumptions are needed to ensure that we can generalize to a new query/domain during testing. A useful technique in the DG literature to generalize to new domains is \textit{domain embedding} \cite{shankar2018generalizing}, which maps domains into the same semantic space. Such a technique captures the domain variations via continuous latent features, and thereby allows effective knowledge transfer between domains.

Along the same line, previous listwise context modeling works \cite{ai2018learning,DCM} choose to construct `context encoding' to capture contextual information using RNNs. Concretely, given the set of feature vectors ${{\left\{ {{\mathbf{h}}_{q,i}} \right\}}_{1\le j\le {{n}_{q}}}}$ for query $q$, these works firstly feed the vectors sequentially into a RNN, and then create the context encoding  ${{\mathbf{c}}_{q}}$ by using the final hidden state of the RNN. As ${{\mathbf{c}}_{q}}$ is then used to query the item vectors to generate the final ranking order, it plays a pivotal role in these models. However, the recurrent nature of RNN means that the creating process of ${{\mathbf{c}}_{q}}$ has no direct access to the candidate items except for the last one, and all relevant information has to be propagated by the RNN through the ordered items in an one-by-one manner, which imposes additional memorization burden on the RNN and may incur unnecessary information loss. 


Based on the above considerations, in the paper we adopt the self-attention mechanism \cite{lin2017structured,vaswani2017attention} to create ${{\mathbf{c}}_{q}}$. Concretely, we first feed ${{\left\{ {{\mathbf{h}}_{q,i}} \right\}}_{1\le i\le {{n}_{q}}}}$ through a multilayer perceptron $\text{ML}{{\text{P}}_{\text{att}}}$ and then a softmax function to generate the attention distribution over the items of the list:
\begin{equation}
\label{attweight}
{{a}_{q,i}}=\frac{\exp \left( \text{ML}{{\text{P}}_{\text{att}}}\left( {{\mathbf{h}}_{q,i}} \right) \right)}{\sum\limits_{k=1}^{{{n}_{q}}}{\exp \left( \text{ML}{{\text{P}}_{\text{att}}}\left( {{\mathbf{h}}_{q,k}} \right) \right)}},1\le i\le n_{q}.
\end{equation}

The generated positive weight ${{a}_{q,i}}$ in (\ref{attweight}) can be interpreted as an estimation of the probability that item ${{d}_{q,i}}$ is the right place to focus on for downstream finer-grained ranking. Based on the attention distribution, we calculate ${{\mathbf{c}}_{q}}$ as the attention-weighted mean of the item vectors:
\begin{equation}
\label{context}
{{\mathbf{c}}_{q}}=\sum\limits_{i=1}^{{{n}_{q}}}{{{a}_{q,i}}{{\mathbf{h}}_{q,i}}}.
\end{equation}
Compared to previous models that simply adopt the last hidden state of RNN, attention layer defined in (\ref{attweight}) and (\ref{context}) may extract a more informative global context encoding ${{\mathbf{c}}_{q}}$ since higher weights are assigned to items that are estimated to be more relevant for context modeling. We then combine ${{\mathbf{c}}_{q}}$ with the original item representations to form the refined representation vectors as:
\begin{equation}
\label{latentcross}
{{\widetilde{\mathbf{h}}}_{q,i}}=\left[ \begin{matrix}
{\mathbf{c}_{q,i}}\odot{{\mathbf{h}}_{q,i}}  \\
{{\mathbf{h}}_{q,i}}  \\
\end{matrix} \right],1\le i\le {{n}_{q}},
\end{equation}
where $\odot$ denotes the Hadamard product. ${{\widetilde{\mathbf{h}}}_{q,i}}$ is regarded as the refined representation since it encodes both the listwise contextual information and the item information that are relevant to the ranking task.

\noindent \textbf{Remark 1.} It is interesting to note that the attention pooling strategy adopted here is a variant of the orderless attention mechanism considered in \cite{OrderlessAttention,lin2017structured}, where all items are independent instead of being sequentially dependent as in RNN-based models. Such a strategy allows the information contained in all item representations to be directly propagated to ${{\mathbf{c}}_{q}}$, which may prevent the long-term memorization problem in RNN-based models. 

\subsection{Query Normalization Layer}

As mentioned earlier, the feature distributions of candidate items for different queries are often different, which makes the learning of an effective global ranking function difficult. This problem cannot be completely solved by adopting the context representation presented in the previous section, as the ranking model still needs to learn to disentangle the information about the `global' differences between queries and the `local' differences between items for each query, and then focus on the latter to perform effective ranking. 

To address the distributional differences among queries, we borrow ideas from previous DG approaches \cite{ganin2016domain,motiian2017unified,muandet2013domain}, and encourage the ranking system to learn \textit{query-invariant} latent representations that have similar or even identical distributions across all queries, with the hope that the system may generalize better to unseen queries by eliminating the detrimental influence of inter-query variability.

Additionally, compared with general DG problems, LETOR has its own distinct characteristics that can be exploited to facilitate the learning of query-invariant representations. In particular, the candidate items for each query is available as a whole, which makes the problem of characterizing the underling distributions easier. For example, for any query $q$, we can directly compute the attention-weighted per-dimension mean and variance vectors of its associated feature distribution as: 
\begin{equation}
\label{crosscontext}
{{\overline{\mathbf{c}}}_{q}}=\sum\limits_{i=1}^{{{n}_{q}}}{{{a}_{q,i}}{{\widetilde{\mathbf{h}}}_{q,i}}},
\end{equation}
\begin{equation}
\label{latentcrossstd}
\overline{\boldsymbol{\upsigma}}_{q}^{2}=\sum\limits_{i=1}^{{{n}_{q}}}{{{a}_{q,i}}{{\left( {{\widetilde{\mathbf{h}}}_{q,i}}-{{\overline{\mathbf{c}}}_{q}} \right)}^{2}}}.
\end{equation}
As is the case with the context vector (\ref{context}), the attention weights (\ref{attweight}) are adopted here to emphasize items that are estimated to be more important. Based on (\ref{crosscontext}) and (\ref{latentcrossstd}), we can construct a query normalization (QN) layer that directly matches these two feature distribution statistics of any given query to zero and unit vectors, respectively:
\begin{equation} 
\label{latentnormal} 
{{\overline{\mathbf{h}}}_{q,i}}=\left( {{\widetilde{\mathbf{h}}}_{q,i}}-{{\overline{\mathbf{c}}}_{q}} \right)\odot {{\left( {{\overline{\boldsymbol{\upsigma}}}_{q}}+\varepsilon  \right)}^{-1}},1\le i\le n_{q},
\end{equation}
where $\varepsilon$ is a small positive constant to avoid division by zero. 

The QN transform (\ref{latentnormal}) can be viewed as a straightforward extension of the widely used batch normalization (BN) transform \cite{ioffe2015batch}, where the key difference is that the latter applies the normalization to a batch of training queries instead of to items in each single query.

\subsection{Ranking Layer}

The normalized hidden representations (\ref{latentnormal}) are passed into another MLP with softmax function to infer the final ranking score: 
\begin{equation}
\label{rankingscore}
{{s}_{q,i}}=\frac{\exp \left( \text{ML}{{\text{P}}_{\text{rank}}}\left( {{\overline{\mathbf{h}}}_{q,i}} \right) \right)}{\sum\limits_{i=1}^{{{n}_{q}}}{\exp \left( \text{ML}{{\text{P}}_{\text{rank}}}\left( {{\overline{\mathbf{h}}}_{q,i}} \right) \right)}},1\le i\le {{n}_{q}}.
\end{equation}
Note that (\ref{attweight}) and (\ref{rankingscore}) have the same structure, thus the ranking layer is essentially the second attention layer in our model, and ${{\left\{ {{s}_{q,i}} \right\}}_{1\le i\le {{n}_{q}}}}$ are the calculated attention weights.  

\subsection{Objective Function}

In general, for each training batch $\mathcal{B}\subseteq \mathcal{T}$, QILCM jointly minimizes the \textit{ranking loss} and the \textit{query confusion loss} with a weight parameter $\lambda$:
\begin{equation}
\label{lemonade}
{{\mathcal{L}}_{\text{QILCM}}}={{\mathcal{L}}_{\text{rank}}}+\lambda {{\mathcal{L}}_{\text{conf}}}.
\end{equation}
In (\ref{lemonade}), the ranking loss $ {\mathcal{L}}_{\text{rank}} $ can be any suitable loss function that measures the ranking accuracy. In this work, we specifically adopt the AttRank loss proposed in \cite{ai2018learning} due to its good performance. Let the relevance labels be normalized as:
\begin{equation}
\label{labelnormmalization}
{{\widetilde{y}}_{q,i}}=\frac{\psi \left( {{y}_{q,i}} \right)}{\sum\limits_{j=1}^{{{n}_{q}}}{\psi \left( {{y}_{q,j}} \right)}},1\le i\le n_{q},
\end{equation}
where $\psi \left( x  \right)$ is the truncated exponential function that returns $\exp \left( x \right)$ if $x>0$ and 0 otherwise. 
The ranking loss simply measures the cross entropy between the score (\ref{rankingscore}) and the normalized relevancy labels:
\begin{equation}
\label{cross entropy}
{{\mathcal{L}}_{\text{rank}}}=\frac{1}{\left| \mathcal{B} \right|}\sum\limits_{q\in \mathcal{B}}{\left( -\frac{1}{{{n}_{q}}}\sum\limits_{i=1}^{{{n}_{q}}}{{{\widetilde{y}}_{q,i}}\log \left( {{s}_{q,i}} \right)} \right)},
\end{equation}
where $\left| \cdot \right|$ denotes the cardinality of a set.

On the other hand, by using the QN layer, the latent feature distribution for any given query is enforced to have zero mean and unit variance in each dimension. However, other types of distributional differences, such as the differences of covariance patterns, can still be present between feature representations from two queries. The query confusion loss is therefore intended to further promote the alignment of distributions of latent features (\ref{latentnormal}) mapped from different queries. In previous DG works, this goal is typically achieved either by minimizing a metric between distributions, or through domain adversarial learning \cite{ganin2016domain,cohen2018cross} which updates the model parameters to fool a jointly learned domain discriminator that attempts to distinguish between samples from different domains. Due to the large number of domains/queries involved, it is difficult to train a domain discriminator in our setting, and thus we focus on the former approach and choose to minimize the following loss:
\begin{equation}
{{\mathcal{L}}_{\text{conf}}}=\frac{1}{{{\left| \mathcal{B} \right|}^{2}}}\sum\limits_{{{q}_{1}}\in \mathcal{B}}{\sum\limits_{{{q}_{2}}\in \mathcal{B}}{{{d}_{\text{CH}}}\left( {{q}_{1}},{{q}_{2}} \right)}},
\end{equation}
where ${{d}_{\text{CH}}}$ is the Chamfer (pseudo)-distance (CD) \cite{fan2017point} for measuring the distance between two point sets:
\begin{equation}
\begin{aligned}
{{d}_{\text{CH}}}\left( {{q}_{1}},{{q}_{2}} \right)&=\sum\limits_{i=1}^{{{n}_{{{q}_{1}}}}}{\underset{1\le j\le {{n}_{{{q}_{2}}}}}{\mathop{\min }}\,\left( \left\| {{\overline{\mathbf{h}}}_{{{q}_{1}},i}}-{{\overline{\mathbf{h}}}_{{{q}_{2}},j}} \right\|_{2}^{2} \right)} \\ 
& +\sum\limits_{j=1}^{{{n}_{{{q}_{2}}}}}{\underset{1\le i\le {{n}_{{{q}_{1}}}}}{\mathop{\min }}\,\left( \left\| {{\overline{\mathbf{h}}}_{{{q}_{1}},i}}-{{\overline{\mathbf{h}}}_{{{q}_{2}},j}} \right\|_{2}^{2} \right)}  
\end{aligned}
\end{equation}
\noindent \textbf{Remark 2.} As both the QN layer and the query confusion regularization are intended to reduce the distributional differences of queries, a naturally arising question is whether these techniques would lead to loss of useful query information and hurt the ranking performance. We will address this issue in the experiments.

\section{Experiments}

\subsection{Baseline Methods}

We compare the proposed QILCM with the following three benchmark methods whose implementations are publicly available:

DCM\footnote{https://www.dropbox.com/s/swghso88q0s3hp7/code.zip} \cite{DCM} and  DLCM\footnote{https://github.com/QingyaoAi/Deep-Listwise-Context-Model-for-Ranking-Refinement} \cite{ai2018learning} are two recently proposed listwise context modeling methods based on RNNs. Note that the initial formulation of DCM could only handle binary-valued relevancy labels, we resolve this problem by simply replacing the original softmax loss function of DCM with the AttRank loss (\ref{rankingscore}) used in DLCM and our method.

LambdaMART \cite{000278621400005} is one of most widely-used listwise LETOR method. We adopt the open-source implementation of this algorithm provided in \cite{capannini2016quality}.

\subsection{Ranking Tasks and Datasets}

We evaluate various methods on two ranking tasks which listwise context modeling methods have been successfully applied to:
\subsubsection{Ranking Refinement.} \cite{ai2018learning} show that the initial ranked list returned by a LETOR method can be greatly improved by using listwise context modeling methods to rerank the top-ranked results. For this task, we used two large-scale LETOR datasets: Istella-S\footnote{http://quickrank.isti.cnr.it/istella-dataset/} \cite{istella} and Microsoft Letor 30K\footnote{https://www.microsoft.com/en-us/research/project/mslr/} \cite{qin2013introducing}. We followed the experimental protocols described in \cite{ai2018learning}, and adopted LambdaMART to do the initial retrieval, items ranked among the top-100 positions were then re-ranked using various listwise context modeling methods. Note that item lists with less than 100 items were entirely re-ranked in our experiments.

\subsubsection{User Choice Ranking.} \cite{DCM} show that listwise context modeling methods can accurately predict the user's choice among alternative commodities. For this task, we used Airline Itinerary\footnote{https://www.dropbox.com/s/qzyt0xwn4u2a1ed/data\_full\_anonym.zip}, which is an anonymized version of the dataset used in \cite{DCM}, each record in the data corresponds to a query, and contains the candidate items presented to a customer. The items that the customers purchased are positive samples and others are negative samples. 

Statistics of the used datasets are summarized in Table \ref{tab:data}.
\begin{table}[htbp]
	\normalsize
		\centering
		\begin{tabular}{c|cccc}
			\hline
			\ Dataset   & Queries   & Items & Rel. & Feats. \\\hline
			\multirow{1}*{Airline Itinerary}
			& 33951  & 1,089k & 2 & 17\\
			\multirow{1}*{Microsoft 30k}
			& 31531  & 3,771k & 5 & 136\\
			\multirow{1}*{Istella-S}
			& 33018 & 3,302k & 5 &  220\\\hline
		\end{tabular}		
	\caption{Characteristics of the datasets used in the experiments: number of queries, items, relevance levels, and features.}
	\label{tab:data}
\end{table}

\subsection{Evaluation Metrics}

For each task, we adopt the evaluation metrics used in prior work. For ranking refinement tasks, we follow \cite{ai2018learning} and use the standard discounted cumulative gain (NDCG) metrics that include NDCG@1, NDCG@3, NDCG@5, and NDCG@10. For Microsoft 30k dataset, an additional metric NDCG@50 was also evaluated. On the other hand, for user choice ranking tasks where relevance feedbacks are binary-valued, we follow \cite{DCM} and use top precision metrics that include P@1(Precision@1) and P@5.

\subsection{Model Training}

Each dataset is split in train, validation and test sets according to a 60\%-20\%-20\% scheme. The validation set was used to select the optimal hyperparameters for all involved methods. We did not perform extensive hyperparameter search for the proposed model, and used virtually the same architecture throughout all the experiments and datasets. More specifically, the dimensions of the nonlinear transformations (\ref{abstract}) in the Input Encoder were fixed as 100, while MLPs used in (\ref{attweight}) and (\ref{rankingscore}) consist of 2 hidden layers with either 256 or 128 ELUs. The models were trained with the Adam algorithm \cite{kingma2014adam} with a learning rate of 0.001, batch size of 80. Training generally converged after less than 100 passes through the entire training dataset.

\subsection{Comparisons with Baseline Methods}

\begin{table*}[tb]
	\footnotesize
	\centering
	\begin{tabular}{c|p{16mm}<{\centering}|p{16mm}<{\centering}p{16mm}<{\centering}p{16mm}<{\centering}p{16mm}<{\centering}p{16mm}<{\centering}p{17mm}<{\centering}}
		\hline
		\ Dataset  & Metrics & QILCM & DCM & DLCM & LambdaMART & Improv. & P-value \\\hline
		\multirow{2}*{Airline Itinerary}
		& P@1 & *0.2833   & \underline{0.2618} & 0.2562 & 0.2327 & 8.21\% & $2.40\times {{10}^{-23}}$ \\
		& P@5 & *0.6958   & 0.6586 & \underline{0.6651} & 0.6239 & 4.61\% & $1.22\times {{10}^{-20}}$ \\\hline
		\multirow{5}*{Microsoft 30k}
		& NDCG@1 & *0.5447   & 0.4938 & \underline{0.4973} & 0.4800 & 9.53\% & $3.83\times {{10}^{-19}}$\\
		& NDCG@3 & *0.5313   & \underline{0.4827} & 0.4811 & 0.4766 & 10.06\% & $7.58\times {{10}^{-16}}$\\
		& NDCG@5 & *0.5368    & \underline{0.4904} & 0.4892 & 0.4842 & 9.46\% & $1.60\times {{10}^{-16}}$\\
		& NDCG@10 & *0.5564   & 0.5093 & \underline{0.5135} & 0.5061 &8.35\% &  $2.27\times {{10}^{-15}}$\\
		& NDCG@50 & *0.6482   & 0.6127 & \underline{0.6146} & 0.6092 &5.46\% &  $4.51\times {{10}^{-15}}$\\\hline
		\multirow{4}*{Istella-S}
		& NDCG@1 & *0.7023   & 0.6762 & \underline{0.6873} & 0.6644 & 2.18\% & $1.74\times {{10}^{-10}}$\\
		& NDCG@3 & *0.6696   & \underline{0.6552} & 0.6537 & 0.6378 & 2.20\% & $1.99\times {{10}^{-8}}$\\
		& NDCG@5 & *0.6953   & \underline{0.6846} & 0.6831 & 0.6741 & 1.56\% & $4.25\times {{10}^{-10}}$\\
		& NDCG@10 & *0.7645   & 0.7558 & \underline{0.7566} & 0.7456 & 1.04\% & $4.67\times {{10}^{-7}}$\\\hline
	\end{tabular}
	
	\caption{Performance comparison of various methods. The results are averaged over 20 random runs, and the best ones are marked with *. The last two columns show the improvement of QILCM over the best baseline algorithm (highlighted with underline), and the corresponding Student's t-test P-values.}
	\label{baseline}
\end{table*}
The performance comparison results of various methods are reported in Table \ref{baseline}. To eliminate the influence of random initiations, all results are averaged over 20 runs. As is shown in this table, QILCM significantly outperforms all the baselines.

\subsection{Ablation Studies}

To elucidate the contributions of the main components of our system, in this section, we test several variants of the proposed model. The tested implementations include:
\begin{itemize}
	\item \textbf{Variant 1}: Our model with the attention-weighted context encoding (\ref{context}) replaced by simple average pooling of the item encoding vectors. Accordingly, the attention-weighted statistics (\ref{crosscontext}) and (\ref{latentcrossstd}) are replaced with standard mean and variance;
	\item \textbf{Variant 2}: Our model without the domain confusion loss;
	\item \textbf{Variant 3}: Our model without the domain confusion loss and QN layer.	
\end{itemize}
The performance results of different implementations of the proposed methods are shown in Table \ref{ablation}, which shows that QILCM consistently achieves the best performance among all model variants.

\begin{table*}[tb]
	\footnotesize
	\centering
	\centering
	\begin{tabular}{c|p{15mm}<{\centering}|p{15mm}<{\centering}p{15mm}<{\centering}p{15mm}<{\centering}p{15mm}<{\centering}p{15mm}<{\centering}}
		\hline
		\ Dataset  & Metrics & QILCM & Variant 1 & Variant 2 & Variant 3  \\\hline
		\multirow{2}*{Airline Itinerary}
		& P@1 & *0.2833   & 0.2749 & 0.2762 & 0.2694  \\
		& P@5 & *0.6958   & 0.6613 & 0.6803 & 0.6724  \\\hline
		\multirow{5}*{Microsoft 30k}
		& NDCG@1 & *0.5447 & 0.5287 & 0.5359 & 0.5139  \\
		& NDCG@3 & *0.5313 & 0.5127 & 0.5294 & 0.4970  \\
		& NDCG@5 & *0.5368 & 0.5230 & 0.5347 & 0.5030  \\
		& NDCG@10 & *0.5564 & 0.5459 & 0.5538 & 0.5229  \\
		& NDCG@50 & *0.6482 & 0.6363 & 0.6438 & 0.6331  \\\hline
		\multirow{4}*{Istella-S}
		& NDCG@1 & *0.7023   & 0.6966 & 0.6982 & 0.6837  \\
		& NDCG@3 & *0.6696   & 0.6654 & 0.6672 & 0.6628  \\
		& NDCG@5 & *0.6953   & 0.6936 & 0.6931 & 0.6923  \\		
		& NDCG@10 & *0.7645   & 0.7622 & 0.7637 & 0.7602  \\\hline
	\end{tabular}
	
	\caption{Performance comparison of different implementations of QILCM. The results are averaged over 20 random runs, and the best ones are marked with *.}
	\label{ablation}
\end{table*}

\subsection{Anomalous Query Analysis}

\begin{figure}
	\centering
	\includegraphics[width=.7\columnwidth]{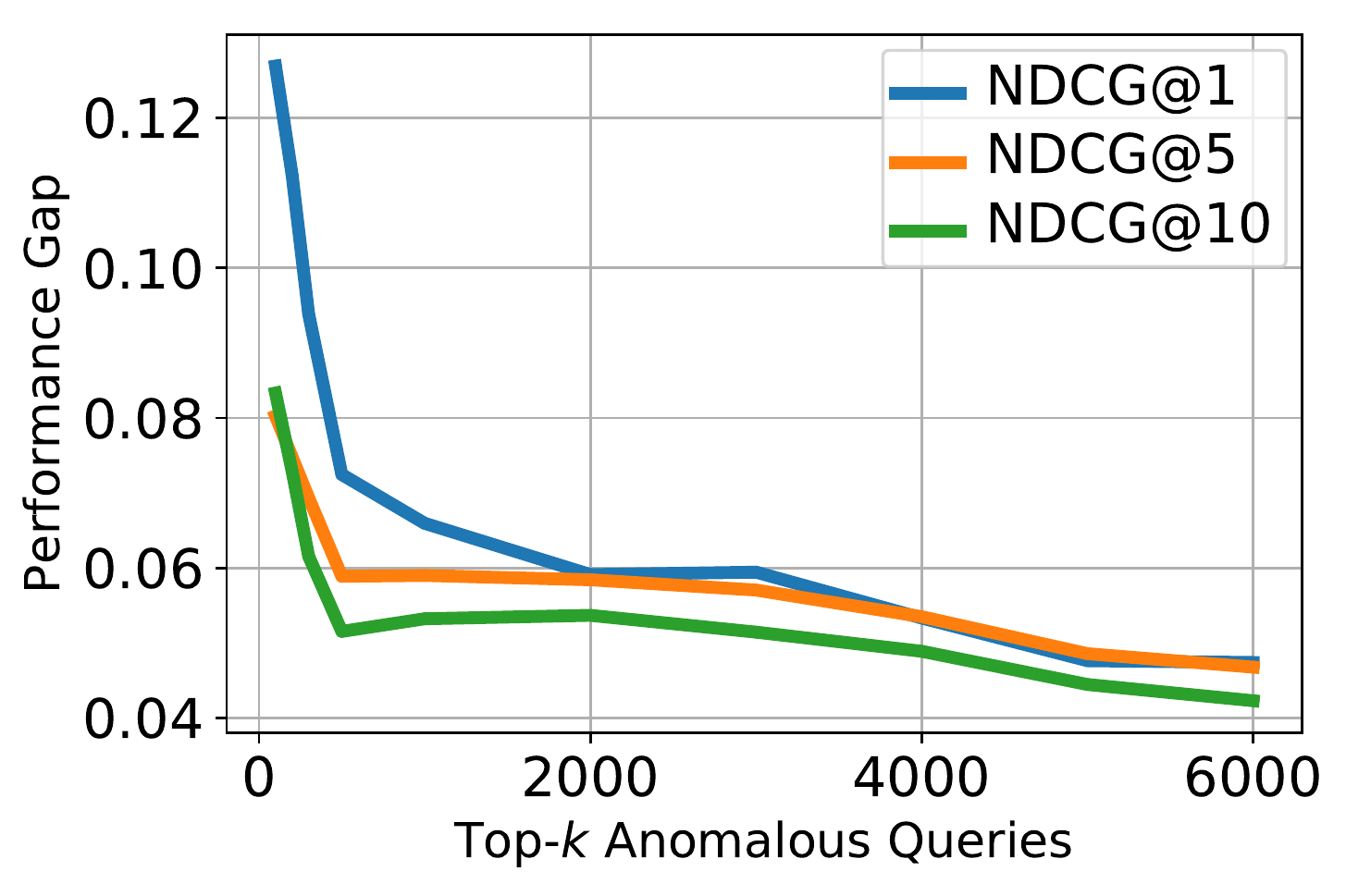} 
	\caption{Performance gap between QILCM and the best-performing baseline (DLCM) on Microsoft 30k dataset.}
	\label{PerfGap}
\end{figure}

To further investigate the performance of QILCM, in this section we conducted additional analysis of experimental results on the Microsoft 30k dataset. Concretely, we firstly followed \cite{geng2008query}, and constructed vector representation for each query by averaging over the feature values of the top \textit{k} ranked items (\textit{k} was set as 10 during the experiments). After that, we fitted Isolation Forest \cite{liu2008isolation} to the training queries, and then used the learned model to assign a `anomaly score' to each query in the test set. Intuitively, this score quantifies how different a query is from the majority of training data, and we examine the performance difference between QILCM and the best-performing baseline (DLCM) for queries with different levels of anomaly. As shown in Figure \ref{PerfGap}, the performance gap between QILCM and DLCM is significantly widened for more anomalous queries. For example, the NDCG@1 gap between QILCM and DLCM is increased from 0.047 to 0.121, which clearly demonstrates the advantage of tackling heterogeneous queries using the proposed DG perspective.

\section{Conclusion}

In this paper, we introduce a DG formulation of the LETOR problem and propose a novel neural architecture for DG in this LETOR context. We evaluate our techniques on three benchmark datasets, demonstrating that the proposed approach outperforms previous state-of-the-art approaches by a substantial margin. 


\bibliographystyle{aaai}
\bibliography{sigproc}

\begin{thebibliography}{}

\bibitem[\protect\citeauthoryear{Ai \bgroup et al\mbox.\egroup
  }{2018}]{ai2018learning}
Ai, Q.; Bi, K.; Guo, J.; and Croft, W.~B.
\newblock 2018.
\newblock Learning a deep listwise context model for ranking refinement.
\newblock In {\em SIGIR},  135--144.

\bibitem[\protect\citeauthoryear{Blanchard, Lee, and
  Scott}{2011}]{blanchard2011generalizing}
Blanchard, G.; Lee, G.; and Scott, C.
\newblock 2011.
\newblock Generalizing from several related classification tasks to a new
  unlabeled sample.
\newblock In {\em NIPS},  2178--2186.

\bibitem[\protect\citeauthoryear{Burges \bgroup et al\mbox.\egroup
  }{2005}]{burges2005learning}
Burges, C.; Shaked, T.; Renshaw, E.; Lazier, A.; Deeds, M.; Hamilton, N.; and
  Hullender, G.
\newblock 2005.
\newblock Learning to rank using gradient descent.
\newblock In {\em ICML},  89--96.

\bibitem[\protect\citeauthoryear{Cao \bgroup et al\mbox.\egroup
  }{2007}]{cao2007learning}
Cao, Z.; Qin, T.; Liu, T.-Y.; Tsai, M.-F.; and Li, H.
\newblock 2007.
\newblock Learning to rank: from pairwise approach to listwise approach.
\newblock In {\em ICML},  129--136.

\bibitem[\protect\citeauthoryear{Capannini \bgroup et al\mbox.\egroup
  }{2016}]{capannini2016quality}
Capannini, G.; Lucchese, C.; Nardini, F.~M.; Orlando, S.; Perego, R.; and
  Tonellotto, N.
\newblock 2016.
\newblock Quality versus efficiency in document scoring with learning-to-rank
  models.
\newblock {\em Information Processing and Management} 52(6):1161--1177.

\bibitem[\protect\citeauthoryear{Chapelle and Chang}{2011}]{YahooL2R}
Chapelle, O., and Chang, Y.
\newblock 2011.
\newblock Yahoo! learning to rank challenge overview.
\newblock In {\em Proceedings of the Learning to Rank Challenge},  1--24.

\bibitem[\protect\citeauthoryear{Chen \bgroup et al\mbox.\egroup
  }{2018}]{chen2018sequential}
Chen, X.; Xu, H.; Zhang, Y.; Tang, J.; Cao, Y.; Qin, Z.; and Zha, H.
\newblock 2018.
\newblock Sequential recommendation with user memory networks.
\newblock In {\em WSDM},  108--116.

\bibitem[\protect\citeauthoryear{Clevert, Unterthiner, and
  Hochreiter}{2015}]{elu}
Clevert, D.-A.; Unterthiner, T.; and Hochreiter, S.
\newblock 2015.
\newblock Fast and accurate deep network learning by exponential linear units
  (elus).
\newblock {\em arXiv preprint arXiv:1511.07289}.

\bibitem[\protect\citeauthoryear{Cohen \bgroup et al\mbox.\egroup
  }{2018}]{cohen2018cross}
Cohen, D.; Mitra, B.; Hofmann, K.; and Croft, W.~B.
\newblock 2018.
\newblock Cross domain regularization for neural ranking models using
  adversarial learning.
\newblock In {\em SIGIR},  1025--1028.

\bibitem[\protect\citeauthoryear{Covington, Adams, and
  Sargin}{2016}]{covington2016deep}
Covington, P.; Adams, J.; and Sargin, E.
\newblock 2016.
\newblock Deep neural networks for youtube recommendations.
\newblock In {\em RecSys},  191--198.

\bibitem[\protect\citeauthoryear{Fan \bgroup et al\mbox.\egroup
  }{2018}]{fan2018modeling}
Fan, Y.; Guo, J.; Lan, Y.; Xu, J.; Zhai, C.; and Cheng, X.
\newblock 2018.
\newblock Modeling diverse relevance patterns in ad-hoc retrieval.
\newblock In {\em SIGIR},  375--384.

\bibitem[\protect\citeauthoryear{Fan, Su, and Guibas}{2017}]{fan2017point}
Fan, H.; Su, H.; and Guibas, L.
\newblock 2017.
\newblock A point set generation network for 3d object reconstruction from a
  single image.
\newblock In {\em CVPR},  2463--2471.

\bibitem[\protect\citeauthoryear{Freno}{2017}]{freno2017practical}
Freno, A.
\newblock 2017.
\newblock Practical lessons from developing a large-scale recommender system at
  zalando.
\newblock In {\em RecSys},  251--259.

\bibitem[\protect\citeauthoryear{Freund \bgroup et al\mbox.\egroup
  }{2003}]{freund2003efficient}
Freund, Y.; Iyer, R.; Schapire, R.~E.; and Singer, Y.
\newblock 2003.
\newblock An efficient boosting algorithm for combining preferences.
\newblock {\em JMLR} 4(Nov):933--969.

\bibitem[\protect\citeauthoryear{Ganin \bgroup et al\mbox.\egroup
  }{2016}]{ganin2016domain}
Ganin, Y.; Ustinova, E.; Ajakan, H.; Germain, P.; Larochelle, H.; Laviolette,
  F.; Marchand, M.; and Lempitsky, V.
\newblock 2016.
\newblock Domain-adversarial training of neural networks.
\newblock {\em JMLR} 17(1):2096--2030.

\bibitem[\protect\citeauthoryear{Gehring \bgroup et al\mbox.\egroup
  }{2017}]{gehring2017convolutional}
Gehring, J.; Auli, M.; Grangier, D.; Yarats, D.; and Dauphin, Y.~N.
\newblock 2017.
\newblock Convolutional sequence to sequence learning.
\newblock In {\em ICML},  1243--1252.

\bibitem[\protect\citeauthoryear{Geng \bgroup et al\mbox.\egroup
  }{2008}]{geng2008query}
Geng, X.; Liu, T.-Y.; Qin, T.; Arnold, A.; Li, H.; and Shum, H.-Y.
\newblock 2008.
\newblock Query dependent ranking using k-nearest neighbor.
\newblock In {\em SIGIR},  115--122.

\bibitem[\protect\citeauthoryear{Ioffe and Szegedy}{2015}]{ioffe2015batch}
Ioffe, S., and Szegedy, C.
\newblock 2015.
\newblock Batch normalization: Accelerating deep network training by reducing
  internal covariate shift.
\newblock In {\em ICML},  448--456.

\bibitem[\protect\citeauthoryear{Karmaker~Santu, Sondhi, and
  Zhai}{2017}]{karmaker2017application}
Karmaker~Santu, S.~K.; Sondhi, P.; and Zhai, C.
\newblock 2017.
\newblock On application of learning to rank for e-commerce search.
\newblock In {\em SIGIR},  475--484.

\bibitem[\protect\citeauthoryear{Kingma and Ba}{2014}]{kingma2014adam}
Kingma, D.~P., and Ba, J.
\newblock 2014.
\newblock Adam: A method for stochastic optimization.
\newblock {\em arXiv preprint arXiv:1412.6980}.

\bibitem[\protect\citeauthoryear{LeCun \bgroup et al\mbox.\egroup
  }{2012}]{lecun2012efficient}
LeCun, Y.~A.; Bottou, L.; Orr, G.~B.; and M{\"u}ller, K.-R.
\newblock 2012.
\newblock Efficient backprop.
\newblock In {\em Neural networks: Tricks of the trade}. Springer.
\newblock  9--48.

\bibitem[\protect\citeauthoryear{Li \bgroup et al\mbox.\egroup
  }{2018}]{li2018learning}
Li, D.; Yang, Y.; Song, Y.-Z.; and Hospedales, T.~M.
\newblock 2018.
\newblock Learning to generalize: Meta-learning for domain generalization.
\newblock In {\em AAAI}.

\bibitem[\protect\citeauthoryear{Li, Wu, and Burges}{2008}]{li2008mcrank}
Li, P.; Wu, Q.; and Burges, C.~J.
\newblock 2008.
\newblock Mcrank: Learning to rank using multiple classification and gradient
  boosting.
\newblock In {\em NIPS},  897--904.

\bibitem[\protect\citeauthoryear{Lin \bgroup et al\mbox.\egroup
  }{2017}]{lin2017structured}
Lin, Z.; Feng, M.; Santos, C. N.~d.; Yu, M.; Xiang, B.; Zhou, B.; and Bengio,
  Y.
\newblock 2017.
\newblock A structured self-attentive sentence embedding.
\newblock In {\em ICLR}.

\bibitem[\protect\citeauthoryear{Liu, Ting, and Zhou}{2008}]{liu2008isolation}
Liu, F.~T.; Ting, K.~M.; and Zhou, Z.-H.
\newblock 2008.
\newblock Isolation forest.
\newblock In {\em ICDM},  413--422.

\bibitem[\protect\citeauthoryear{Lucchese \bgroup et al\mbox.\egroup
  }{2016}]{istella}
Lucchese, C.; Nardini, F.~M.; Orlando, S.; Perego, R.; Silvestri, F.; and
  Trani, S.
\newblock 2016.
\newblock Post-learning optimization of tree ensembles for efficient ranking.
\newblock In {\em SIGIR},  949--952.

\bibitem[\protect\citeauthoryear{Manotumruksa, Macdonald, and
  Ounis}{2018}]{manotumruksa2018contextual}
Manotumruksa, J.; Macdonald, C.; and Ounis, I.
\newblock 2018.
\newblock A contextual attention recurrent architecture for context-aware venue
  recommendation.
\newblock In {\em SIGIR},  555--564.

\bibitem[\protect\citeauthoryear{Motiian \bgroup et al\mbox.\egroup
  }{2017}]{motiian2017unified}
Motiian, S.; Piccirilli, M.; Adjeroh, D.~A.; and Doretto, G.
\newblock 2017.
\newblock Unified deep supervised domain adaptation and generalization.
\newblock In {\em ICCV},  5716--5726.

\bibitem[\protect\citeauthoryear{Mottini and Acuna-Agost}{2017}]{DCM}
Mottini, A., and Acuna-Agost, R.
\newblock 2017.
\newblock Deep choice model using pointer networks for airline itinerary
  prediction.
\newblock In {\em SIGKDD},  1575--1583.

\bibitem[\protect\citeauthoryear{Muandet, Balduzzi, and
  Sch{\"o}lkopf}{2013}]{muandet2013domain}
Muandet, K.; Balduzzi, D.; and Sch{\"o}lkopf, B.
\newblock 2013.
\newblock Domain generalization via invariant feature representation.
\newblock In {\em ICML},  10--18.

\bibitem[\protect\citeauthoryear{Nallapati}{2004}]{nallapati2004discriminative}
Nallapati, R.
\newblock 2004.
\newblock Discriminative models for information retrieval.
\newblock In {\em SIGIR},  64--71.

\bibitem[\protect\citeauthoryear{Qin and Liu}{2013}]{qin2013introducing}
Qin, T., and Liu, T.-Y.
\newblock 2013.
\newblock Introducing letor 4.0 datasets.
\newblock {\em arXiv preprint arXiv:1306.2597}.

\bibitem[\protect\citeauthoryear{Raffel and Ellis}{2015}]{OrderlessAttention}
Raffel, C., and Ellis, D.~P.
\newblock 2015.
\newblock Feed-forward networks with attention can solve some long-term memory
  problems.
\newblock In {\em ICLR}.

\bibitem[\protect\citeauthoryear{Shankar \bgroup et al\mbox.\egroup
  }{2018}]{shankar2018generalizing}
Shankar, S.; Piratla, V.; Chakrabarti, S.; Chaudhuri, S.; Jyothi, P.; and
  Sarawagi, S.
\newblock 2018.
\newblock Generalizing across domains via cross-gradient training.
\newblock In {\em ICLR}.

\bibitem[\protect\citeauthoryear{Taylor \bgroup et al\mbox.\egroup
  }{2008}]{SoftRank}
Taylor, M.; Guiver, J.; Robertson, S.; and Minka, T.
\newblock 2008.
\newblock Softrank: Optimizing non-smooth rank metrics.
\newblock In {\em WSDM},  77 -- 85.

\bibitem[\protect\citeauthoryear{Vaswani \bgroup et al\mbox.\egroup
  }{2017}]{vaswani2017attention}
Vaswani, A.; Shazeer, N.; Parmar, N.; Uszkoreit, J.; Jones, L.; Gomez, A.~N.;
  Kaiser, {\L}.; and Polosukhin, I.
\newblock 2017.
\newblock Attention is all you need.
\newblock In {\em NIPS},  5998--6008.

\bibitem[\protect\citeauthoryear{Vinyals, Bengio, and
  Kudlur}{2016}]{OrderMatters}
Vinyals, O.; Bengio, S.; and Kudlur, M.
\newblock 2016.
\newblock Order matters: Sequence to sequence for sets.
\newblock In {\em ICLR}.

\bibitem[\protect\citeauthoryear{Volkovs and
  Zemel}{2012}]{volkovs2012collaborative}
Volkovs, M., and Zemel, R.~S.
\newblock 2012.
\newblock Collaborative ranking with 17 parameters.
\newblock In {\em NIPS},  2294--2302.

\bibitem[\protect\citeauthoryear{Wu \bgroup et al\mbox.\egroup
  }{2010}]{000278621400005}
Wu, Q.; Burges, C. J.~C.; Svore, K.~M.; and Gao, J.
\newblock {2010}.
\newblock {Adapting boosting for information retrieval measures}.
\newblock {\em {Information Retrieval}} {13}({3}):{254--270}.

\bibitem[\protect\citeauthoryear{Wu, Hsieh, and Sharpnack}{2018}]{sqlrank}
Wu, L.; Hsieh, C.-J.; and Sharpnack, J.
\newblock 2018.
\newblock {SQL}-rank: A listwise approach to collaborative ranking.
\newblock In {\em ICML},  5315--5324.

\bibitem[\protect\citeauthoryear{Xiang \bgroup et al\mbox.\egroup
  }{2010}]{xiang2010context}
Xiang, B.; Jiang, D.; Pei, J.; Sun, X.; Chen, E.; and Li, H.
\newblock 2010.
\newblock Context-aware ranking in web search.
\newblock In {\em SIGIR},  451--458.

\bibitem[\protect\citeauthoryear{Xie \bgroup et al\mbox.\egroup
  }{2017}]{xie2017controllable}
Xie, Q.; Dai, Z.; Du, Y.; Hovy, E.; and Neubig, G.
\newblock 2017.
\newblock Controllable invariance through adversarial feature learning.
\newblock In {\em NIPS},  585--596.

\bibitem[\protect\citeauthoryear{Zamani \bgroup et al\mbox.\egroup
  }{2017}]{zamani2017situational}
Zamani, H.; Bendersky, M.; Wang, X.; and Zhang, M.
\newblock 2017.
\newblock Situational context for ranking in personal search.
\newblock In {\em WWW},  1531--1540.

\bibitem[\protect\citeauthoryear{Zamani \bgroup et al\mbox.\egroup
  }{2018}]{zamani2018neural}
Zamani, H.; Mitra, B.; Song, X.; Craswell, N.; and Tiwary, S.
\newblock 2018.
\newblock Neural ranking models with multiple document fields.
\newblock In {\em WSDM},  700--708.

\bibitem[\protect\citeauthoryear{Zhou \bgroup et al\mbox.\egroup
  }{2018}]{ATRANK}
Zhou, C.; Bai, J.; Song, J.; Liu, X.; Zhao, Z.; Chen, X.; and Gao, J.
\newblock 2018.
\newblock Atrank: An attention-based user behavior modeling framework for
  recommendation.
\newblock In {\em AAAI},  4564--4571.

\bibitem[\protect\citeauthoryear{Zhuang, Ou, and
  Wang}{2018}]{zhuang2018globally}
Zhuang, T.; Ou, W.; and Wang, Z.
\newblock 2018.
\newblock Globally optimized mutual influence aware ranking in e-commerce
  search.
\newblock In {\em IJCAI},  3725--3731.

\end{thebibliography}

\end{document}